# ConnectionChain: Secure Interworking of Blockchains

Shingo Fujimoto
*Security Laboratory*
*FUJITSU Laboratories LTD*
*Kawasaki, Japan*
shingo_fujimoto@fujitsu.com

Yoshiki Higashikado
*Security Laboratory*
*FUJITSU Laboratories LTD*
*Kawasaki, Japan*

Takuma Takeuchi
*Security Laboratory*
*FUJITSU Laboratories LTD*
*Kawasaki, Japan*

*Abstract*—Blockchain is a core technology to manage the value of cryptocurrencies, or to record trails of important business trades. The Smart Contract on blockchain is expected to improve security on blockchain system with automated operation, but it cannot be the solution when the application service required to operate tightly related blockchain ledgers as service business logic. This paper proposed the method to extend the functionality of traditional Smart Contract on blockchain, and introduced the prototype system, named 'ConnectionChain'.

## I. INTRODUCTION

Blockchain is a core technology to manage the values of cryptocurrencies, or to record trails of important business trades, which is categorized as Distributed Ledger Technology, and its application field are getting wider.

The original blockchain, Bitcoin[1] is a management system of cryptocurrency, and its reliability is proven by the fact that Bitcoin is keep running from its born in year 2009 without any interruption. But cryptocurrencies were stolen by criminal attackers according to public news. This could be possible when application service, such as wallet service, is involved. The Smart Contract is expected to improve security on blockchain system against such threats since it can eliminate human involvements with its automation.

Even Smart Contract is ideal solution on improvement for operation at single blockchain ledger, it cannot be the solution when the application service required to operate tightly related blockchain ledgers as service business logic.

This paper proposed the concept of "Extended Smart Contract", which extends the concept of Smart Contract to support automatic operations on tightly related blockchain ledgers. This paper also explains about our prototype system, named ConnectionChain, which is platform service to execute proposed Extended Smart Contract.

## II. BACKGROUNDS

### A. Blockchain technology

Blockchain system is the system consist from P2P(Peer-to-Peer) networked computers, called 'node', which shares single ledger between blockchain participants.

P2P network of blockchain provides high availability by sharing operation on ledger and its operation result, usually called Transaction(abbreviated as Tx in figure), among participating blockchain nodes even when some of nodes are not available.

Blockchain also provides integrity of shared ledger by using cryptography, e.g. secure-hash function, public key encryption, and digital signature as these combination. After each transaction is verified whether it can be accepted or not, acceptable transactions are collected as a group called 'block'. Then, block data is prepared from transactions of the block by calculating hash-value using secure hash function to digest transactions in the block. Note that the input for the secure hash function also includes hash-value of previous block. Since hash-value of previous block digests content of the transactions in previous block(s), new hash-value digests all transactions accepted in the blockchain. That means the integrity of block data is reinforced by adding more number of following block data when time passed.

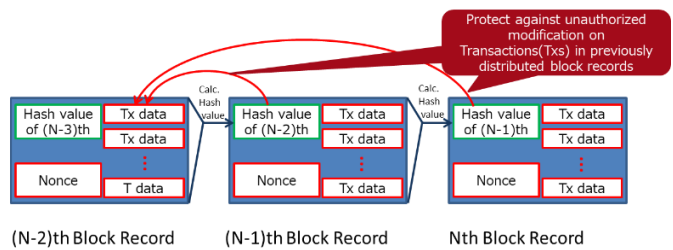

Fig. 1. Fundamental mechanism of blockchain ledger

## B. Smart Contract

Based on the fundamental mechanisms of blockchain, some implementations of blockchain platform equip a security mechanism to execute ledger operation(s) with complexed business logic, called Smart Contract.

Smart Contract is the computer program built on blockchain, and it will be run on the nodes. Smart Contract takes the confirmation and performs series of actions written as the code following rules and policies agreed between parties without the need of oversight from a trusted third party. Smart Contract is expected to shorten the time that the settlement procedure takes by automation. This automation also contribute to avoid dishonest act by system administrators with visible code.

Smart Contract is now considered as disruptive technology for internet society since it enables collaboration between untrusted parties over the internet in de-centralized way.

## III. ISSUES ON USE OF BLOCKCHAIN

Even blockchain is useful technology to manage digitalized asset securely, it requires all involving parties to participate on same blockchain. That is difficult for applying blockchain for real business use cases since the blockchain requires to agree on single rule or policy. Its difficulty can be imaged from the case of making a contract document with multiple parties in real business. Additionally, the performance of processing transactions by blockchain will be getting worse when the number of participating nodes are increased as the nature of P2P network.

Blockchain is sometimes compared with distributed database technology since both provides similar functionality and characteristic. The difference between them is in its administration policy, centralized system or de-centralized system.The best practice to design database is to divide information into small tables to avoid conflict of administrative roles and decreasing performance.

We think this best practice in database can be also applicable for blockchain, if information shared among large number of parties can be divided into smaller ledgers which is managed by smaller administrative parties.

The issue in this approach is how to build the trust on integration of multiple ledgers by security technology.

There are several existing works which archived to use multiple ledgers as an integrated service, such as COSMOS[2] and ILP[3], but they are applicable for limited field with trusting service provider.

## IV. SECURE INTERWORKING BETWEEN BLOCKCHAINS

We are proposing the method to build trust for an integration service working with multiple blockchain ledgers.

We developed a security technology for blockchain platform named 'ConnectionChain', which allows to implement extended functionality of Smart Contract.

### A. Extending functionality of Smart Contract on blockchain

Applicability of Smart Contract on blockchain is limited on executing business logic with condition checks and ledger operations under agreement among the blockchain participants. Our proposed Extend Smart Contract is extended the functionality of Smart Contract to enable executing business logic which requires condition checks and ledger operations crossover multiple blockchain ledgers by written code.

Extended Smart Contract is consist with a runtime environment of Smart Contract on our dedicated blockchain platform and 'interworking node(s)' which will execute remote operation on interworking blockchain and report the result of the operation to the Extended Smart Contract engine. The interworking node acts as normal node from the viewpoint of interworking blockchain participants, and that makes easy to persuade other participants since it will not be required to have any administrative privileges on the blockchain.

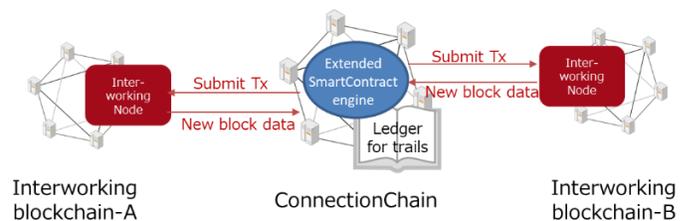

Fig. 2. Extended Smart Contract engine and interworking nodes

The interworking node will issue normal transaction request to interworking blockchain, and report the result of transaction execution. This report is sent from the interworking node with forwarding the block data which is distributed to all nodes on interworking blockchain.

The data format of block data may vary on each blockchain platform implementations, but they are common in having an TXID(Transaction Identifier) which is used to obtain the result of each transaction, and a hash value to check integrity of the transaction result in the interworking blockchain. ConnectionChain considers transaction results in the forwarded blockchain data as trustworthy information for checking condition in business logic of Extended Smart Contract.

The Extended Smart Contract will change internal state on ledger with recording, when it issues a transaction to the interworking blockchain or when it received execution result as the part of verifiable blockchain data. We think this mechanism provides enough visually for acceptable security model of Smart Contract on blockchain.

### B. Escrow Trade on Extended Smart Contract

Extended Smart Contract is required to deal with the case of failure on issued transaction since those transactions are issued without human involved operations. In such case, escrowed feature of Extended Smart Contract is useful. Escrow trade is well-known as trading system for online auctions when the seller and the buyer cannot trust each other. In escrow trade, buyer deposits the money to the system, and the money can be settled only when sold item is delivered to the buyer properly. Extended Smart Contract can execute an 'escrow trade' which execute a transaction on some blockchain, only when execution of depending transaction on another blockchain resulted successfully.

Difficulty to enable escrow trading on blockchain is comes from there are no way to cancel the transaction once it is recorded on the blockchain ledger. Extended Smart Contract solved this with having an account or wallet to hold the deposited asset from the user. When the escrow trade is required, the Extended Smart Contract asks to the user to transfer ownership of the digitalized asset to the Extended Smart Contract, and the Extended Smart Contract monitors the status of the depending transaction on another blockchain. When the depending transaction resulted successful or non-successful, the Extended Smart Contract will transfer ownership of deposited asset to receiver user or sender user as refund depending on the result. Since Extended Smart Contract will be implemented as Smart Contract on a blockchain, honesty on execution of business logic will be guaranteed.

## V. PROTOTYPE IMPLEMENTATION 'CONNECTIONCHAIN'

We implemented a prototype system to prove the concept of Extended Smart Contract, named "ConnectionChain".

### A. Platform service for Extended Smart Contract

ConnectionChain is a platform service which is implemented built on open-sourced blockchain platform, Hyperledger Fabric v1.3 (abbreviated as HLF). HLF is categorized 'consortium chain', and its blockchain will not fork with trust among consortium members.

Core functionality of ConnectionChain is implemented as a chaincode which is Smart Contract in HLF. The chaincode can be written in programing languages such as GO, Java, and JavaScript. The chaincode can use 'WorldState' to share internal state as key-value pair among participating nodes. When a transaction is submitted by HLF client, each node of HLF will execute chaincode to simulate the execution result, and simulated result will be applied on shared ledger including changes on the WorldState using block data only when enough number of participating nodes agreed on the simulated result.

The user of ConnectionChain service can invoke a chaincode which is implemented to perform business logic for an application service. Invoked chaincode of ConnectionChain will issue transaction requests which should be handled by interworking blockchain via the interworking node deployed on the blockchain. ConnectionChain will record transaction ID of issued transaction with timestamp, and starts monitoring the status change receiving block data from the blockchain.

ConnectionChain will determine successful result of submitted transaction request with parsing block data which is usually formatted as JSON (JavaScript Object Notation). ConnectionChain verifies the integrity of each block data by checking if recalculated hash value is matched with the hash value included in it. On the other hand, failure can be determined by the un-successful return code or timeout.

ConnectionChain provides RESTful API to execute Extended Smart Contract, and to monitor progress of the execution.

### B. Use case of ConnectionChain

ConnectionChain is design to support various kind of Extended Smart Contracts, and we will introduce 'payment service for cryptocurrencies' as a sample application.

The user (Alice) of this service will make payment for the commercial item purchase in owning cryptocurrency (A-coin) to the merchant (Bob) who can accept payment in another cryptocurrency (B-coin) only (see Fig.3).

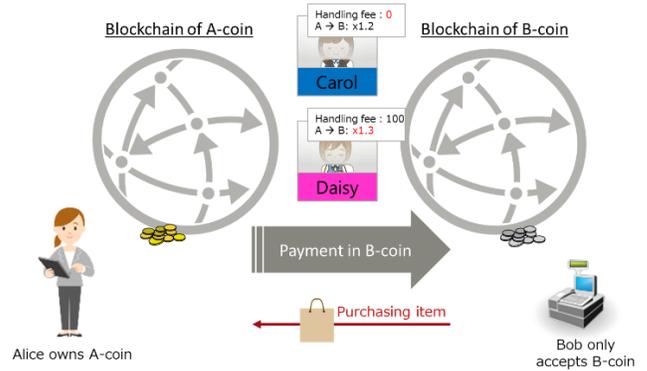

Fig. 3. Sample application 'payment service for cryptocurrencies'

We assumed two service providers, Carol and Daisy, are coexist in the system, and they can configure their own service profile with account numbers of both cryptocurrency blockchain, rate for conversion, and handling fee for settlement.

The use of the service will execute an Extended Smart Contract with the parameters which specifies source account of user, destination account of merchant, selected service profile (Carol's profile), and amount of cryptocurrency in destination chain.

The business logic of this service is implemented to perform following scenario:

1. Initialize internal state of Extended Smart Contract on ConnectionChain for the settlement.

2. ConnectionChain issues an asset transfer request from Alice to the account of ConnectionChain at A-coin.

3. Block data of A-coin which contains the successful result of the asset transfer is notified to ConnectionChain.

4. ConnectionChain verified integrity of the data, and it prepares an asset transfer request from account of exchanger Charlie's account at B-coin to Bob's account at B-coin with applying Charlie's profile in conversion.

5. ConnectionChain checks if balance of Charlie's account at B-coin is enough to execute the prepared request.

6. ConnectionChain issues prepared request to blockchain of B-coin.

7. Block data of B-coin which contains the result for execution the execution of prepared request is notified to ConnectionChain.

8. ConnectionChain verified integrity of the data, and it will issue to transfer deposited A-coin from user Alice to exchanger Charlie's account at A-coin when it detects successful result of the prepared request, or it will issue to transfer deposited A-coin to Alice as refund.

9. Block data of A-coin which contains the successful result of transfer request issued by ConnectionChain in A-coin.

10. ConnectionChain acknowledges the end of process.

Fig.4 shows the execution result in successful case, and Fig.5 shows the execution result in un-successful case.

Fig. 4. Result of Extended Smart Contract (successful)

Fig. 5. Result of Extended Smart Contract (un-successful)

## VI. CONCLUSION

We proposed the method to extend the functionality of traditional Smart Contract on blockchain, and explained prototype implementation named 'ConnectionChain'.

We also introduced 'cryptocurrencies settlement service' as one of possible use cases of ConnectionChain.

Since use of blockchain technology is spreading from financial field to other area rapidly, we want to enhance our prototype implementation to support various type of blockchains.

REFERENCES

[1] S. Nakamoto, "Bitcoin: A Peer-to-Peer Electronic Cash System", Nov. 2008.[Online] https://bitcoin.org/bitcoin.pdf .

[2] J. Kwon and B. Buchman, "COSMOS white paper", Jan. 2019 [Online] https://github.com/cosmos/cosmos/blob/master/WHITEPAPER.md

[3] S. Thomas and E. Schwartz, "A Protocol for Interledger Payments", Jul. 2016.[Online] https://interledger.org/interledger.pdf

[4] M.Herlihy, "Atomic Cross-Chain Swaps.", Jul. 2018, Proceedings of the 2018 ACM Symposium on Principles of Distributed Computing, Pages 245-254

[5] L. Kan, et al "A Multiple Blockchains Architecture on Inter-Blockchain Communication," 2018 IEEE International Conference on Software Quality, Reliability and Security Companion (QRS-C), Lisbon, 2018, pp. 139-145.

[6] A.Hope-Bailie and S. Thomas, "Interledger: Creating a Standard for Payments", Apr. 2016, the 25th International Conference Companion

[7] Hyperledger Quilt [Online], https://www.hyperledger.org/projects/quilt. [Accessed:22-Jul-2019].